# Transitions between liquid crystalline phases investigated by dielectric and infra-red spectroscopies

Aleksandra Deptuch[1,*], Natalia Osiecka-Drewniak[1], Anna Paliga[2], Natalia Górska[3], Anna Drzewicz[1], Katarzyna Chat[1], Mirosława D. Ossowska-Chruściel[4], Janusz Chruściel[4]

[1] Institute of Nuclear Physics Polish Academy of Sciences, Radzikowskiego 152, PL-31342 Kraków, Poland

[2] Faculty of Physics, Astronomy and Applied Computer Science, Jagiellonian University, Łojasiewicza 11, PL-30348 Kraków, Poland

[3] Faculty of Chemistry, Jagiellonian University, Gronostajowa 2, PL-30387 Kraków, Poland

[4] Faculty of Science, University of Siedlce, 3 Maja 54, PL-08110 Siedlce, Poland

* corresponding author, aleksandra.deptuch@ifj.edu.pl

**Abstract**

The liquid crystalline 11OS5 compound, forming the nematic phase and a few smectic phases, is investigated by broadband dielectric spectroscopy and infra-red spectroscopy. The dielectric relaxation times, ionic conductivity, and positions of infra-red absorption bands corresponding to selected intra-molecular vibrations are determined as a function of temperature in the range from isotropic liquid to a crystal phase. The correlation coefficient matrix and k-means cluster analysis of infra-red spectra are tested for detection of phase transitions. The density-functional theory calculations are carried out for interpretation of experimental infra-red spectra. The performance of various basis sets and exchange-correlation functionals is compared, including both agreement of scaled calculated band positions with experimental values and computational time. The inter-molecular interactions in the crystal phase are inferred from the experimental IR spectra and density-functional theory calculations for dimers in head-to-head and head-to-tail configurations. The experimental temperature dependence of the C=O stretching band suggests that the head-to-tail configuration in a crystal phase is more likely. A significant slowing down of the flip-flop relaxation process is observed at the transition between the smectic C and hexagonal smectic X phases.

**1. Introduction**

The broadband dielectric spectroscopy (BDS) gives insight into the relaxation processes of dipoles in a weak, alternating electric field [1]. In liquid crystals, it involves molecular processes – rotations of molecules around their short and long axes and, in some cases, also collective processes – fluctuations of order parameters [2]. The BDS method is applied for identification of liquid crystalline phases [3-5], determination of phase transition temperatures [5-7], as well as investigation of the glass transition [5,8] and crystallization kinetics [8,9]. The complementary infra-red (IR) spectroscopy reveals the intra-molecular vibrations which change a molecule's dipole moment [10]. This method is applied for investigation of the sample's purity [11,12], molecular conformation [13,14], and intermolecular interactions [14,15]. Some IR absorption bands are sensitive to phase transitions, especially these related to significant changes in structure, like melting of a crystal [14,15].

The purpose of this study is analysis of the dielectric and IR spectra of 4-pentylphenyl-4'-undecyloxythiobenzoate, denoted as 11OS5 (Figure 1), focused on transitions between the mesomorphic and crystal phases of this compound. 11OS5 forms several thermotropic liquid crystalline phases between isotropic liquid (Iso) and crystal (Cr) [16]:

- Nematic (N) with a long-range orientational order of long molecular axes [17].
- Smectic A (SmA) with a quasi-long-range lamellar order and zero average tilt of molecules within smectic layers [17].
- Smectic C (SmC) with quasi-long-range lamellar order and non-zero average tilt of molecules within smectic layers [17]. The tilt angle in the SmC phase of 11OS5, estimated from the X-ray diffraction patterns, reaches only 11° [16].
- Smectic X (SmX), a crystal-like tilted smectic phase with a long-range positional order described by a monoclinic unit cell and a hexagonal intra-layer order. It is either the SmG phase with the unit cell parameters $b < a < c$, $\alpha = \gamma = 90°$, $\beta \neq 90°$ or the SmJ phase with $a < b < c$, $\alpha = \gamma = 90°$, $\beta \neq 90°$ [17]. The X-ray diffraction data from [16] do not enable distinction between these phases, which differ only by the direction of a molecular tilt in respect to a hexagonal ordering within layers [17]. The tilt angle of molecules in the SmX phase of 11OS5 is equal to ca. 15° [16].
- Smectic Y' and Y, crystal-like smectic phases, with a long-range positional order described by the monoclinic unit cells and probably a herring-bone intra-layer order [17]. The X-ray diffraction patterns of these phases were not obtained due to quick crystallization [16].

The phase sequence of 11OS5 during cooling at 5 K/min is Iso (357.4 K) N (355.9 K) SmA {338} SmC (326.0 K) SmX (298.5 K) SmY' (296.0 K) SmY (294.7 K) Cr, where temperatures in () were determined by differential scanning calorimetry and temperature in {} was determined by polarizing optical microscopy, because the SmA → SmC transition was not visible in the calorimetric results [16]. The SmX, SmY', SmY phases are metastable and they are observed only after supercooling below the melting temperature of a crystal, which is 334 K [16]. However, 11OS5 remains in the metastable SmX phase long enough to enable conducting measurements in isothermal conditions before crystallization occurs [16].

In the first part, the relaxation times, dielectric strength, and ionic conductivity are obtained as a function of temperature from the BDS spectra of 11OS5 collected on cooling. In the second part, the experimental IR spectrum of 11OS5 in a crystal phase at the room temperature is analyzed based on theoretical spectra calculated at various levels of theory. The experimental IR absorption bands are assigned to particular intra-molecular vibrations and performances of different levels of theory (scaling coefficients, root-mean square error, computational times) are compared. In the third part, the IR spectra of 11OS5 collected on cooling are analyzed using the matrix of correlation coefficients [18] and k-means cluster method [19-21]. Positions of selected IR absorption bands are investigated as a function of temperature. The main aim is an attempt to detect subtle changes in the BDS and IR spectra related to the SmA → SmC transition as well as to investigate more significant changes related to the SmC → SmX transition. The comparison of BDS and IR data may enable correlation of dielectric relaxation with intra-molecular vibrations for better understanding of molecular rearrangements during phase transitions between mesophases and a crystalline state.

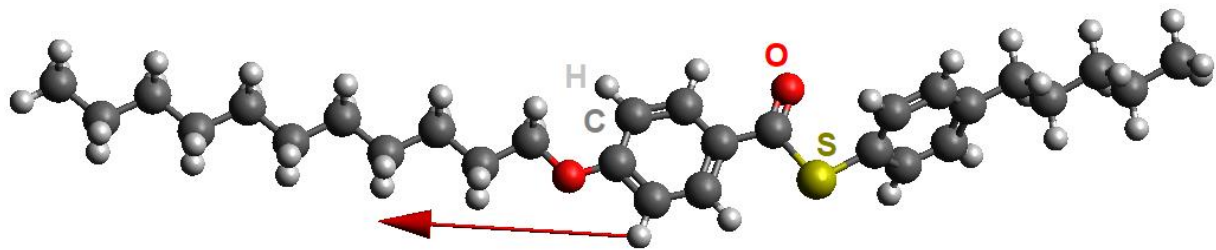

Figure 1. The 11OS5 molecular model and its dipole moment (2.4 D) optimized by the DFT method (def2TZVPP basis set, B3LYP-D3(BJ) functional).

## 2. Experimental and computational details

The synthesis of 4-pentylphenyl-4'-undecyloxythiobenzoate, abbreviated as 11OS5, is described elsewhere [22]. The IR results presented in this work confirm the molecular structure.

The BDS measurements were performed using the Novocontrol Technologies spectrometer for a sample with a thickness of 75 μm, placed between two golden electrodes with the polytetrafluoroethylene spacer. The dielectric spectra were collected on cooling from 373 K to 273 K, in the frequency range of $0.1$-$10^7$ Hz. The data were analyzed by fitting the complex model of the dielectric permittivity vs. frequency in OriginPro.

The Fourier-transform IR measurements were performed using the Bruker VERTEX 70v FT-IR spectrometer equipped with Advanced Research System DE-202A cryostat and ARS-2HW compressor. The IR spectra were collected by the attenuated total reflection method for a pure sample at the room temperature in the wavenumber range of 370-4000 $cm^{-1}$ and by the transmission method for a sample mixed with KBr on cooling from 373 K to 273 K in the wavenumber range of 450-4000 $cm^{-1}$, both with a resolution of 2 $cm^{-1}$.

The correlation coefficient matrix of IR spectra collected in different temperatures was calculated using a Python script based on the SciPy and NumPy libraries [23]. The fitting of pseudo-Voigt peak functions with a linear background to selected absorption bands and k-means cluster analysis with the Euclidean metrics [19] were carried out in OriginPro. The background subtraction was performed before calculation of the correlation coefficient matrix and k-means cluster analysis.

The assignment of the IR absorption bands was based on the theoretical IR spectra calculated by the DFT method at various levels of theory for an isolated molecule in Gaussian 16, Revision C.01 [24]. The tested basis sets were def2SVP, def2SVPP, def2TZVP, def2TZVPP [25,26], 631+Gd, and 6311+Gdp [27-30]. The applied exchange-correlation functionals were either BLYP-D3(BJ) or B3LYP-D3(BJ) [31-33]. The contributions of vibrational modes were calculated via the potential energy distribution analysis in VEDA [34,35].

## 3. Results and discussion
### 3.1. Dielectric spectra vs. temperature

The BDS spectra of 11OS5 (Figure 2) were fitted with the complex function including the Cole-Cole model of relaxation processes, contribution from conductivity in the imaginary part $\varepsilon''$, and electrode polarization contribution in the real part $\varepsilon'$ of dielectric permittivity $\varepsilon^*$ [2,36]:

$$\varepsilon^*(f) = \varepsilon'(f) - i\varepsilon''(f) = \varepsilon_\infty + \sum_j \frac{\Delta \varepsilon_j}{1+(2\pi i f \tau_j)^{1-a_j}} - \frac{is_1}{(2\pi f)^{n_1}} + \frac{is_2}{(2\pi f)^{n_2}}. \qquad (1)$$

The Cole-Cole model describes each relaxation processes by the dielectric strength $\Delta\varepsilon$, relaxation time $\tau$, and parameter $a$, corresponding to the distribution of the relaxation time (for the Debye model, $a = 1$); $\varepsilon_\infty$ is the dielectric dispersion in the limit of high frequencies; $s_1$, $n_1$ and $s_2$, $n_2$ are fitting parameters of the background at low frequencies, originating from the sample's conductivity and electrode polarization, respectively, or from the tails of low-frequency relaxation processes. The BDS results are summarized in Figure 3. The phase transitions are visible as significant discontinuities in the temperature dependence of the dielectric dispersion registered at ca. 1 kHz. The exception is the SmA → SmC transition, which corresponds to a very small step in the $\varepsilon'(1\text{ kHz}, T)$ plot. The hexagonal SmX, despite it is metastable, is observed in a temperature range wider than 15 K, while the SmY', SmY phases are not detected and the direct SmX → Cr transition occurs.

Two relaxation processes are observed in the isotropic liquid, nematic, SmA and SmC phases. The fitting results of Equation (1) are presented only for temperatures below the Iso → N transition, because in the isotropic liquid, some of the fitting parameters had too large uncertainties. The low-frequency process, with the average $\Delta\varepsilon = 5.9$, $a = 0.11$ and activation energy $E_a = 79(2)$ kJ/mol, is interpreted as the Maxwell-Wagner-Sillars process [37]. The high-frequency process, with the average $\Delta\varepsilon = 0.11$, $a = 0.01$ and activation energy $E_a = 94.2(4)$ kJ/mol, is the molecular s-process, originating from rotations of molecules around their short axes (called also flip-flop motions) [6,38,39]. The presented activation energies were obtained in the SmA and SmC phases. For the N phase, which has a narrow temperature range, only two BDS spectra were collected and the activation energies were not determined. The conductivity contribution in the low-frequency region of $\varepsilon''$ can be fitted with an assumption of ionic conductivity, where $n_1 = 1$ [40]. The conductivity determined as $\sigma = s_1 \varepsilon_0$, where $\varepsilon_0$ is the vacuum permittivity, shows the linear dependence in the activation plot, with $E_a = 97(3)$ kJ/mol. The polarization contribution in the nematic phase is described by $n_2 = 1.5$, in agreement with [40], while in SmA and SmC, this contribution was fixed to zero. The activation energies of the s-process obtained earlier for the SmA and SmC phases in nOS5 homologs with n = 8, 9, 10 are in the 92-110 kJ/mol range [41], comparable to the result for 11OS5.

In the hexagonal SmX phase, the s-process shifts abruptly to lower frequencies. The ratio of $\tau$ in the SmX and SmC phases at the SmC → SmX transition is equal to 46. The dielectric strength of the s-process decreases to $\Delta\varepsilon = 0.07$ and the activation energy increases to 144(1) kJ/mol. The observed changes are caused by arise of the positional order within the smectic layers. The Cole-Cole parameter equals on average $a = 0.04$, thus, the s-process still has an almost ideally Debye character. The conductivity part cannot be correctly described under an assumption of $n_1 = 1$. The fitted values of $n_1$ are equal to 0.8-0.9 and only shortly before crystallization, $n_1 = 0.4$. The activation energy of $\sigma$, which is only an effective conductivity if $n_1 \neq 1$, is equal to 99(2) kJ/mol. Thus, the SmC → SmX transition leads to decrease in conductivity, but $E_a$ is the same within uncertainties. The polarization contribution also is not anymore described by $n_2 = 1.5$, as it was in the nematic phase. Instead, $n_2$ gradually decreases on cooling from 1.1 to 0.6. Deviation from $n_2 = 1.5$ may be caused by the Maxwell-Wagner relaxation, which is out of the measured frequency range, but the high-frequency tail still affects the collected spectra.

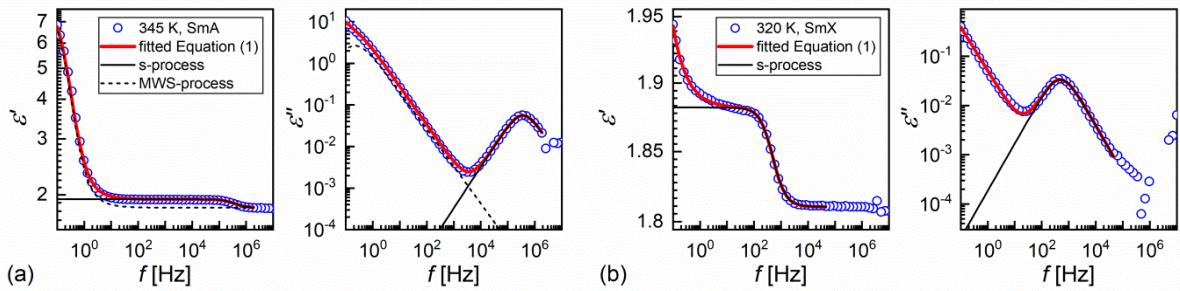

Figure 2. Exemplary BDS spectra of 11OS5 in the SmA (a) and SmX (b) phases with the fitting results of Equation (1).

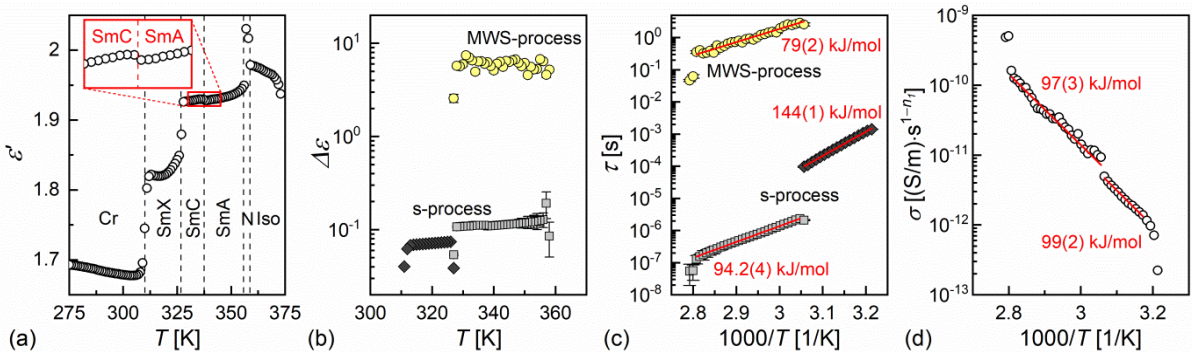

Figure 3. Dielectric dispersion at 1 kHz (a) and dielectric strength of relaxation processes in 11OS5 (b) as a function of temperature; activation plots of the relaxation times (c) and conductivity (d).

*3.2. Scaling of calculated infra-red spectra*

The assignment of IR absorption bands to intra-molecular vibrations was performed for the IR spectrum collected in the reflection mode at the room temperature for 11OS5 in a crystal phase (Figure 4). The theoretical band positions and intensities were calculated using the triple-zeta basis sets def2TZVPP, def2TZVP; split-valence basis sets def2SVPP, def2SVP; and 631+Gd, 6311+Gdp basis sets with diffuse corrections. The Becke exchange functional and Lee-Yang-Parr correlation functional were applied as BLYP and B3LYP functionals together with the Grimme's 3D dispersion and Becke-Johnson damping (see Section 2 for references). The calculated spectra are shown in Figure 4 below the experimental result. The band assignments as well as the comparison of experimental and calculated band positions are gathered in Tables S1-S8 in the Supplementary Materials (SM).

The scaling coefficient $a$ relates experimental $\tilde{v}_{exp}$ and calculated $\tilde{v}_{calc}$ band positions as follows: $\tilde{v}_{exp} = a\tilde{v}_{calc}$. It is determined as a slope of the linear fit to the $\tilde{v}_{exp}(\tilde{v}_{calc})$ plot, with the intercept fixed to zero (Figure S1 in SM). The fitting can be performed in the full wavenumber range. However, it is advised to determine it separately for the <1000 cm$^{-1}$, 1000-2000 cm$^{-1}$, and >2000 cm$^{-1}$ ranges [42]. The results for 11OS5 (Table 1) show that the scaling coefficients at the same level of theory are larger and closer to 1 for the bands located below 2000 cm$^{-1}$ than for these at higher wavenumbers. The scaling coefficients obtained for <1000 cm$^{-1}$ and 1000-2000 cm$^{-1}$ have close values, in some cases even equal within the uncertainty. In all cases where the B3LYP-D3(BJ) functional was applied, $a < 1$, which means that calculated band positions are overestimated. The BLYP-D3(BJ) functional leads to underestimation of band positions only below 2000 cm$^{-1}$, $a > 1$, while the linear fit performed at higher wavenumbers and in the whole range gives $a < 1$.

The root-mean square error between the experimental and scaled calculated band positions was obtained as $RMSE = \sqrt{\frac{1}{n}\sum_{i=1}^{n}(\tilde{v}_{expi} - a\tilde{v}_{calci})^2}$, where $n$ is the number of bands. Different values of $a$ were used in the <1000 cm$^{-1}$, 1000-2000 cm$^{-1}$, and >2000 cm$^{-1}$ ranges, according to Table 1. The overall scaling factor (column 'all' in Table 1) was not used in calculation of $RMSE$. The smallest $RMSE$ values are 11.6 cm$^{-1}$ for def2TZVPP/B3LYP-D3(BJ) and 11.9 cm$^{-1}$ for def2TZVP/B3LYP-D3(BJ). The corresponding computational time is 156.9 h and 85.8 h, respectively. Thus, the longest def2TZVPP/B3LYP-D3(BJ) calculations do not give a significant improvement in $RMSE$ compared to twice as short def2TZVP/B3LYP-D3(BJ) calculations. The third smallest $RMSE$, 12.5 cm$^{-1}$, is obtained for 631+Gd/B3LYP-D3(BJ), with a relatively short computational time, 17.6 h. Interestingly, much longer 6311+Gd/B3LYP-D3(BJ) calculations lasting 40.7 h give slightly worse $RMSE$ equal to 13.1 cm$^{-1}$. One of the shortest are def2SVP/BLYP-D3(BJ) calculations, lasting 4.7 h, and the corresponding $RMSE$ is 13.6 cm$^{-1}$. The BLYP-D3(BJ) functional shows a worse performance when used with the 631+Gd basis set, because the computational time is longer (8.8 h) and $RMSE$ is larger (15.1 cm$^{-1}$) than for the def2SVP basis set. The largest $RMSE$ values are 17.1 cm$^{-1}$ for

def2SVPP/B3LYP-D3(BJ) and 17.8 cm$^{-1}$ def2SVP/B3LYP-D3(BJ), with the computational times 4.7 h and 6.6 h, respectively.

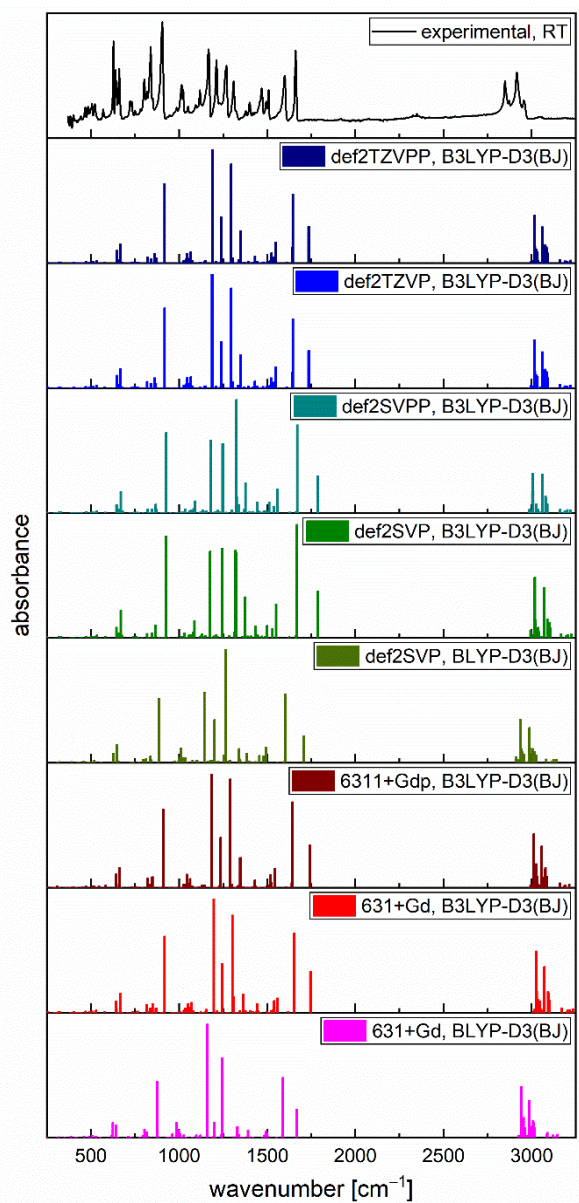

Figure 4. Experimental IR spectrum of 11OS5 in a crystal phase at the room temperature compared to unscaled spectra calculated by the DFT method at different levels of theory.

Table 1. Scaling factors between experimental and calculated IR band positions determined for 11OS5. The first column refers to the detailed IR band assignments shown in SM. The root mean square error ($RMSE$) was calculated for experimental and scaled calculated band positions using the scaling factors obtained separately for the <1000 cm$^{-1}$, 1000-2000 cm$^{-1}$, and >2000 cm$^{-1}$ ranges. The cpu times were taken from the Gaussian output files.

| table | basis set | functional | scaling factor | | | | RMSE [cm$^{-1}$] | cpu time [h] |
|---|---|---|---|---|---|---|---|---|
| | | | <1000 cm$^{-1}$ | 1000-2000 cm$^{-1}$ | >2000 cm$^{-1}$ | all | | |
| S1 | def2TZVPP | B3LYP-D3(BJ) | 0.974(4) | 0.974(2) | 0.952(2) | 0.961(2) | 11.6 | 156.9 |
| S2 | def2TZVP | B3LYP-D3(BJ) | 0.975(4) | 0.974(2) | 0.952(2) | 0.962(2) | 11.9 | 85.8 |
| S3 | def2SVPP | B3LYP-D3(BJ) | 0.966(5) | 0.960(4) | 0.952(2) | 0.952(2) | 17.1 | 4.7 |
| S4 | def2SVP | B3LYP-D3(BJ) | 0.969(5) | 0.965(4) | 0.950(2) | 0.957(2) | 17.8 | 6.6 |
| S5 | def2SVP | BLYP-D3(BJ) | 1.005(3) | 1.003(3) | 0.975(2) | 0.988(3) | 13.6 | 4.7 |
| S6 | 6311+Gdp | B3LYP-D3(BJ) | 0.978(4) | 0.975(3) | 0.953(2) | 0.963(2) | 13.1 | 40.7 |
| S7 | 631+Gd | B3LYP-D3(BJ) | 0.977(4) | 0.966(3) | 0.947(2) | 0.957(2) | 12.5 | 17.6 |
| S8 | 631+Gd | BLYP-D3(BJ) | 1.014(4) | 1.004(4) | 0.974(2) | 0.988(3) | 15.1 | 8.8 |

*3.3 Infra-red spectra vs. temperature*

The IR spectra of 11OS5 were collected on cooling from 373 K to 273 K in the transmission mode (Figure 5). The spectra in the isotropic liquid, nematic, and smectic phases are very similar and only crystallization shows as sharpening of absorption bands. Three methods are applied to search for the subtle differences in IR spectra in higher temperatures: (1) calculation of the correlation coefficients matrix, (2) k-means cluster analysis, and (3) determination of the selected absorption bands' positions as a function of temperature.

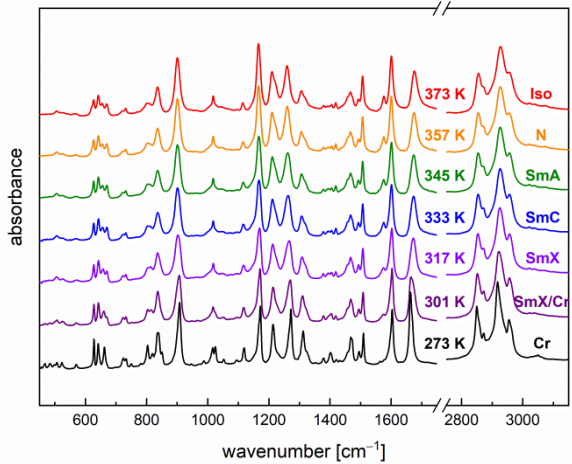

Figure 5. Representative experimental IR spectra of 11OS5 collected on cooling.

The correlation coefficient of two datasets is defined as [18]:

$$r = \frac{\sum_{i=1}^{n}(x_i-\bar{x})(y_i-\bar{y})}{(n-1)s_x s_y}, \qquad (2)$$

where $x_i$, $y_i$ are values obtained at different temperatures, $\bar{x}$, $\bar{y}$ are mean values, $s_x$, $s_y$ are standard deviations, $n$ is a number of $(x_i, y_i)$ pairs. The matrix of the correlation coefficients of the IR spectra of 11OS5 is presented in Figure 6a. The data can be divided into four clusters: {373, 369, 365, 357 K}

– Iso and N; {353, 349, 345, 341, 337, 333, 329 K} – SmA and SmC; {325, 321, 317, 313, 309, 305 K} – SmX; 301 K – crystallization; {297, 293, 289, 285, 281, 277, 273 K} – crystal. The correlation coefficients for some temperatures (e.g. 317 K) enable distinction of the N phase at 357 K (Figure 6b). The SmA → SmC transition does not show in the correlation coefficient matrix.

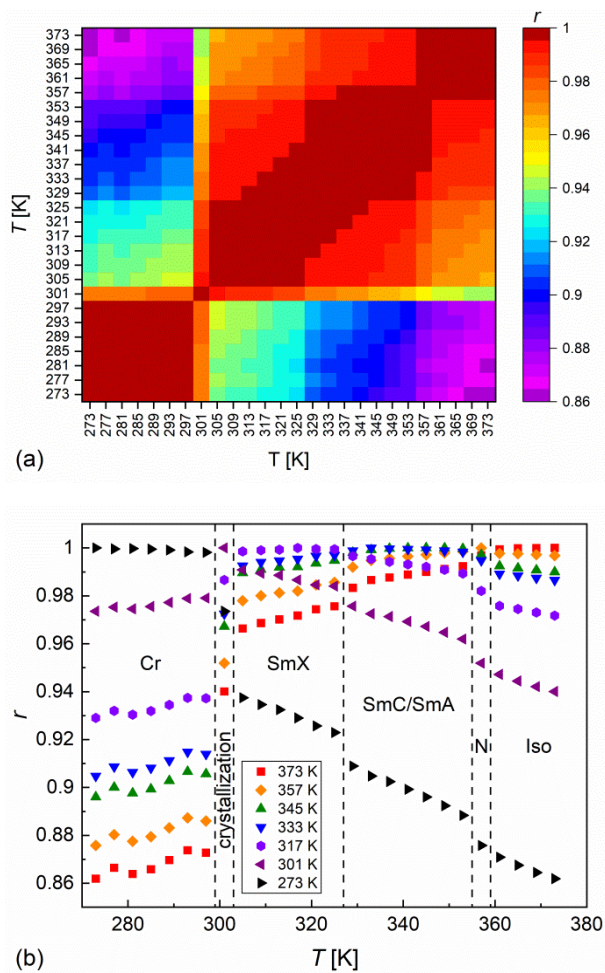

Figure 6. Correlation coefficient matrix of IR spectra of 11OS5 (a) and correlation coefficients for selected spectra as a function of temperature (b).

The k-means cluster analysis is based on an algorithm which separates data into clusters with maximal similarity of data within each cluster and maximal difference of data between clusters [19-21]. The optimal number of clusters indicated by the "elbow method" [20] is four (Figure S2 in SM). The true number of clusters in the dataset is unknown. The elbow method is a heuristic technique used to estimate the optimal number of clusters by identifying the point where increasing the number of clusters no longer significantly reduces the within-cluster variance. The 4-cluster analysis performed for IR spectra in the whole range divides data as follows: {373, 369, 365, 357 K} – Iso and N; {353, 349, 345, 341, 337, 333, 329, 325, 321 K} – SmA, SmC, and SmX; {317, 313, 309, 305, 301 K} – SmX and crystallization; {297, 293, 289, 285, 281, 277, 273 K} – crystal. Thus, the 4-cluster

analysis for full spectra is sensitive to the N → SmA and SmX → Cr transitions, but does not separate correctly the temperature ranges of SmC and SmX. The Iso and N phase are not distinguished, same as SmA and SmC. In the next step, the 4-cluster analysis was performed separately for selected spectral ranges to find out which intra-molecular vibrations are sensitive to particular phase transitions:

- 450-590 cm$^{-1}$ – out-of-plane deformations of both aromatic rings, in-plane deformations of the aromatic ring next to the alkoxy chain, scissoring of C-S-C, scissoring of C-C-C and C-O-C in the alkoxy chain,
- 590-685 cm$^{-1}$ – out-of-plane deformations of the aromatic ring next to the alkoxy chain, in-plane deformations of both aromatic rings, out-of-plane deformations of the COS group,
- 685-760 cm$^{-1}$ – out-of-plane deformations of the aromatic ring next to the alkyl chain, rocking of CH$_2$ groups in both terminal chains,
- 760-860 cm$^{-1}$ – out-of-plane deformations of both aromatic rings, in-plane deformations of the aromatic ring next to the alkoxy chain, rocking of CH$_2$ groups in the alkoxy chain, twisting of CH$_2$ groups in the alkyl chain
- 860-930 cm$^{-1}$ – in-plane deformations of the aromatic ring next to the alkoxy chain, C-S stretching in the COS group,
- 930-1065 cm$^{-1}$ – stretching of C-C and C-O bonds in the alkoxy chain,
- 1065-1190 cm$^{-1}$ – in-plane deformations of both aromatic rings, stretching of C-C bonds in the alkoxy chain,
- 1190-1290 cm$^{-1}$ – in-plane deformations of the aromatic ring next to the alkoxy chain, wagging of CH$_2$ groups in the alkyl chain,
- 1290-1345 cm$^{-1}$ – in-plane deformations of both aromatic rings, wagging of CH$_2$ groups in the alkyl chain,
- 1345-1426 cm$^{-1}$ – in-plane deformations of the aromatic ring next to the alkyl chain, wagging of CH$_2$ groups in both terminal chains,
- 1426-1482 cm$^{-1}$ – scissoring of CH$_2$ groups in both terminal chains,
- 1482-1550 cm$^{-1}$ – in-plane deformations of both aromatic rings, scissoring of CH$_2$ groups in the alkoxy chain,
- 1550-1625 cm$^{-1}$ – in-plane deformations of the aromatic ring next to the alkoxy chain,
- 1625-1750 cm$^{-1}$ – stretching of C=O bond,
- 2700-3150 cm$^{-1}$ – stretching of C-H bonds.

The band assignment is based on the def2TZVPP/B3LYP-D3(BJ) calculations in Gaussian [24] and potential energy distribution analysis in VEDA [34,35] (Table S1). The results of 4-cluster analysis identical with these of full IR spectra are obtained for two spectral ranges (Figure 7): 590-685 cm$^{-1}$ (deformations of aromatic rings and out-of-plane deformation of COS) and 1426-1482 cm$^{-1}$ (scissoring of CH$_2$ groups). The absorption bands sensitive to the Iso → N transition in the 4-cluster analysis are these in the 2700-3150 cm$^{-1}$ range (stretching of C-H bonds). The N → SmA transition is indicated by bands at 590-685 cm$^{-1}$, 760-860 cm$^{-1}$, 1065-1190 cm$^{-1}$, and 1290-1750 cm$^{-1}$ (deformations of aromatic rings, out-of-plane deformation of the COS group, rocking, wagging, twisting, and scissoring of CH$_2$ groups, stretching of C-C and C=O bonds). The SmA and SmC phases are separated correctly for bands at 685-760 cm$^{-1}$ (out-of-plane deformations of the aromatic ring next to the alkyl chain, rocking of CH$_2$ groups). The SmC → SmX transition is indicated by bands at 860-930 cm$^{-1}$, 1065-1290 cm$^{-1}$, and 1482-1625 cm$^{-1}$ (deformations of aromatic rings, stretching of C-S in the COS group, stretching of C-C bonds, wagging and scissoring of CH$_2$ groups). The crystallization is detected by all bands. However, the analysis for bands at 450-590 cm$^{-1}$, 685-760 cm$^{-1}$, and 930-1065 cm$^{-1}$ (deformations of aromatic rings, scissoring of C-S-C, C-C-C, C-O-C, stretching of C-C, C-O bonds, rocking of CH$_2$ groups) separates the spectrum collected at 273 K from other spectra collected in the 277-297 K range, which may indicate some structural changes within a solid state.

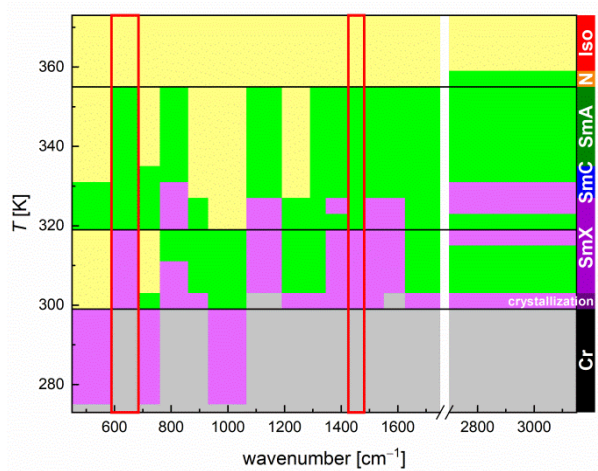

Figure 7. Results of the 4-cluster analysis for selected ranges in the IR spectra of 11OS5. Horizontal lines indicate borders between clusters obtained for full IR spectra. Red frames indicate spectral ranges with identical clusters as obtained for full spectra.

Finally, the temperature behavior of two selected absorption bands is investigated (Figure 8). The first band is the one located at 1663-1674 cm$^{-1}$, corresponding to the C=O stretching ($\nu$C=O). The second band in the one located at 901-909 cm$^{-1}$, corresponding to the in-plane deformations of the aromatic ring close to the alkoxy chain ($\beta_{asym}$Ph$_{alkoxy}$) and C-S stretching in the COS group ($\nu$CS). Each band is well-separated from neighbor absorption bands. The band related to $\nu$C=O shows a redshift (decrease of the band position) at transitions from the less ordered to more ordered phases on cooling, which is explained by formation of weak hydrogen bonds [43,44]. The largest redshift of almost 8 cm$^{-1}$ occurs between the SmX and crystal phases. The redshift at the SmC → SmX transition is very small, less than 1 cm$^{-1}$, and comparable to the redshift observed at the Iso → N → SmA transitions. No redshift occurs at the SmA → SmC transition. The band related to $\beta_{asym}$Ph$_{alkoxy}$ and $\nu$CS vibrations shows the opposite behavior, which is a blueshift (increase of the band position) on cooling. The blueshift at the SmX → Cr and SmC → SmX transitions is 4 cm$^{-1}$ and 1 cm$^{-1}$, respectively, while it is negligible at transitions at higher temperatures.

The influence of intermolecular interactions on IR spectra was modelled by DFT calculations for dimers. The head-to-head and head-to-tail dimers of 11OS5 were optimized (Figure 9) and corresponding IR spectra were calculated using the 631+Gd basis set and B3LYP-D3(BJ) functional (Figure 10), which gave reasonably low *RMSE* and short computational time in calculations for an isolated molecule (Table 1). The theoretical band position at the 631+Gd/B3LYP-D3(BJ) level related to νC=O is 1746 cm$^{-1}$ for an isolated molecule (Table S7 in SM), 1741, 1756 cm$^{-1}$ for a head-to-head dimer, and 1726, 1740 cm$^{-1}$ for a head-to-tail dimer. The theoretical band position at the 631+Gd/B3LYP-D3(BJ) level related to $\beta_{asym}Ph_{alkoxy}$ and νCS is 919 cm$^{-1}$ for an isolated molecule (Table S7 in SM), 912, 922 cm$^{-1}$ for a head-to-head dimer, and 915, 920 cm$^{-1}$ for a head-to-tail dimer. The blueshift of the νC=O band (1756 cm$^{-1}$) occurs when the C=O group is located close to the S atom from a neighbor molecule (C…S distance 4.0 Å) in a head-to-head dimer. The redshift (1726 cm$^{-1}$) occurs when the C=O group interacts with the CH$_2$ groups in the alkoxy chain from a neighbor molecule (O…H distance 2.6 Å) in a head-to-tail dimer. The latter result matches qualitatively the experimental observations. The band related to $\beta_{asym}Ph_{alkoxy}$ and νCS vibrations of the same molecule in a head-to-tail dimer is the weakly blueshifted one at 920 cm$^{-1}$, which also agrees qualitatively with experimental results. Thus, 11OS5 molecules in a crystal phase are likely in a head-to-tail configuration, where C=O groups interact with alkoxy chains. There may be also an intercalated configuration, where C=O groups interact with alkyl chains. The head-to-head configuration is less likely, as it leads to proximity of C=O groups and S atoms from neighbor molecules, which would cause blueshift of the νC=O band, against the experimental results.

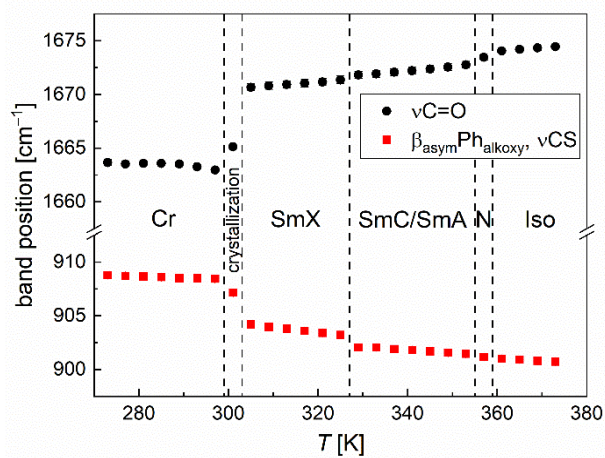

Figure 8. Positions of absorption bands related to νC=O and $\beta_{asym}Ph_{alkoxy}$, νCS vibrations in the 11OS5 molecule vs. temperature.

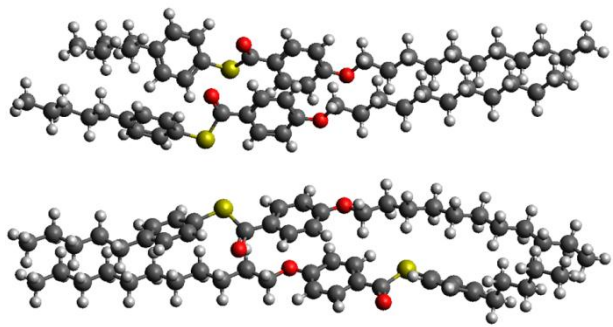

Figure 9. Head-to-head (top) and head-to-tail (bottom) dimers of 11OS5 optimized at the 631+Gd/B3LYP-D3(BJ) level.

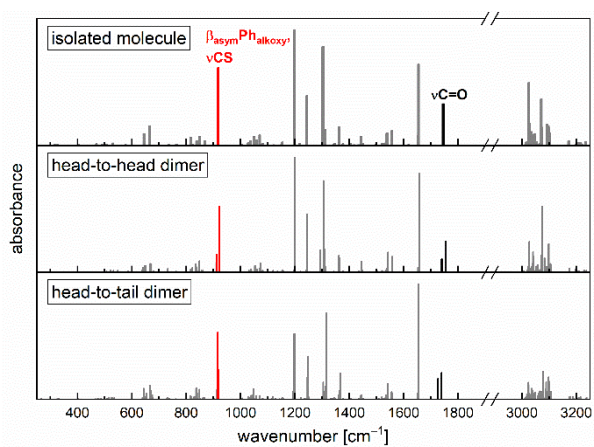

Figure 10. Theoretical unscaled IR spectra calculated for an isolated molecule, head-to-head dimer, and head-to-tail dimer of 11OS5 at the 631+Gd/B3LYP-D3(BJ) level.

**4. Summary and conclusions**

The isotropic liquid, nematic, smectic A, C, X, and crystal phase of the 11OS5 compound were investigated by dielectric and IR spectroscopies. The dielectric dispersion at 1 kHz is sensitive to all phase transition, including the SmC → SmA transition. The dielectric spectra reveal two relaxation processes in the Iso, N, SmA, and SmC phases: low-frequency Maxwell-Wagner-Sillars relaxation and high-frequency molecular s-process, related to rotations around the short molecular axes (flip-flop motion). The MWS process is out of the frequency range in the hexagonal SmX phase. The s-process is also strongly shifted to lower frequencies at the SmC → SmX transition, by more of one order of magnitude. The activation energy of the s-process increases to 144 kJ/mol in SmX compared to 94 kJ/mol in SmA and SmC. No relaxation processes are detected in a crystal phase.

The IR absorption bands were assigned to intra-molecular vibrations based on DFT calculations performed with def2TZVPP, def2TZVP, def2SVPP, def2SVP, 6311+Gdp, 631+Gd basis sets and B3LYP-D3(BJ) or BLYP-D3(BJ) exchange-correlation functionals. The def2TZVPP/B3LYP-D3(BJ) level of theory provides the lowest root-mean square error between experimental and scaled calculated band positions (11.6 cm$^{-1}$). However, def2TZVP/B3LYP-D3(BJ) gives a negligibly higher *RMSE* (11.9 cm$^{-1}$) with twice as short calculations than for def2TZVPP/B3LYP-D3(BJ). The third best performance is obtained for 631+Gd/B3LYP-D3(BJ), which provides $RMSE$ = 12.5 cm$^{-1}$, while computational time is ten times shorter than for def2TZVPP/B3LYP-D3(BJ). The 631+Gd/B3LYP-D3(BJ) level was further applied for calculations of the theoretical spectra of 11OS5 dimers in head-to-head and head-to-tail configurations.

The IR spectra collected as a function of temperature were analyzed by calculation of the correlation coefficient matrix, which was sensitive to the most of phase transitions except the SmC → SmA transition. The k-means cluster analysis performed for the full spectra was sensitive to the N → SmA and SmX → Cr transition, while the Iso → N, SmA → SmC transition were not detected and the SmC → SmX transition temperature was underestimated. The k-means cluster analysis performed separately in various spectral ranges showed that different intra-molecular vibrations had various sensitivity to particular phase transitions. In some cases, IR spectra collected in the same phase were placed in separate clusters by the algorithm. Thus, the prior knowledge of the phase sequence is necessary and the k-means cluster analysis should be used mainly to indicate which absorption bands are the most sensitive to the phase transition in question, at least in a situation where the ordering of molecules occurs gradually on cooling, as it is for liquid crystals.

The absorption band at 1663-1674 cm$^{-1}$, related to the C=O stretching, and the band at 901-909 cm$^{-1}$, related to the in-plane deformations of the aromatic ring close to the alkoxy chain and C-S stretching in the COS group, show the redshift and blueshift with decreasing temperature, respectively. The Iso → N and N → SmA transitions influence mainly the band at 1663-1674 cm$^{-1}$ (redshift of ~1 cm$^{-1}$ at each transition). The IR spectra of SmC and SmA cannot be distinguished by any applied analytical method because the molecular arrangement and, consequently, inter-molecular interactions in these phases are very similar, as the tilt angle in SmC is only 11° and the smectic layer shrinkage between SmA and SmC is ~1% [16]. The redshift/blueshift of the former/latter band at the SmC → SmX transition is only ~1 cm$^{-1}$, despite that the SmX phase is characterized by the long-range positional order in three dimensions. It is caused by the preserved orientational disorder in the SmX phase, detected in dielectric spectra as the s-process. The strongest redshift/blueshift (8/4 cm$^{-1}$) of the former/latter band occurs during crystallization. It agrees with the absence of dielectric relaxation processes in the solid state, indicating lack of the orientational and conformational disorder in a crystal phase of 11OS5. The DFT calculations for dimers indicate that the C=O group interacts rather with the terminal chains than with the S atom, therefore, the head-to-tail or intercalated configuration in a crystal phase is more likely than the head-to-head configuration.

**Acknowledgements.** We thank Assoc. Prof. Ewa Juszyńska-Gałązka from Institute of Nuclear Physics Polish Academy of Sciences for discussion regarding data analysis. We gratefully acknowledge Polish high-performance computing infrastructure PLGrid (HPC Center: ACK Cyfronet AGH) for providing computer facilities and support within computational grant no. PLG/2024/017946. Bruker VERTEX 70v FT-IR spectrometer with Advanced Research System DE-202A cryostat and ARS-2HW compressor were purchased thanks to the European Regional Development Fund in the framework of the Innovative Economy Operational Program (contract no. POIG.02.01.00-12-023/08).

**Declaration statement:** The authors declare no conflict of interest.

# Transitions between liquid crystalline phases investigated by dielectric and infra-red spectroscopies

Aleksandra Deptuch[1,*], Natalia Osiecka-Drewniak[1], Anna Paliga[2], Natalia Górska[3], Anna Drzewicz[1], Katarzyna Chat[1], Mirosława D. Ossowska-Chruściel[4], Janusz Chruściel[4]

[1] Institute of Nuclear Physics Polish Academy of Sciences, Radzikowskiego 152, PL-31342 Kraków, Poland

[2] Faculty of Physics, Astronomy and Applied Computer Science, Jagiellonian University, Łojasiewicza 11, PL-30348 Kraków, Poland

[3] Faculty of Chemistry, Jagiellonian University, Gronostajowa 2, PL-30387 Kraków, Poland

[4] Faculty of Science, University of Siedlce, 3 Maja 54, PL-08110 Siedlce, Poland

* corresponding author, aleksandra.deptuch@ifj.edu.pl

# Supplementary Materials

Table S1. The band assignment of the experimental IR spectra of 11OS5 at the room temperature, based on the DFT calculations (**def2TZVPP** basis set, **B3LYP-D3(BJ)** functional). Notations: β – in-plane deformation, γ – out-of-plane deformation, δ – scissoring, ν – stretching, ρ – rocking, τ – twisting, ω – wagging.

| experimental peak position [cm$^{-1}$] | unscaled calculated peak position [cm$^{-1}$] | scaled calculated peak position [cm$^{-1}$] | vibration (contribution ≥ 10%) |
|---|---|---|---|
| 384 | 387 | 377 | 58% δCCC$_{alkyl}$ |
| 401 | 403 | 393 | 27% δCCC$_{alkoxy}$ |
| 416 | 417 | 406 | 83% γPh$_{alkyl}$ |
| 445 | 448 | 436 | 20% δCCO$_{alkoxy}$ |
| 468 | 472 | 460 | 40% δCCC$_{alkoxy}$ |
| 484 | 495 | 482 | 12% γPh$_{alkyl}$ |
| 500 | 505 | 492 | 40% δCCC$_{alkyl}$ |
| 506 | 515 | 502 | 73% γPh$_{alkoxy}$ |
| 523 | 532 | 518 | 49% δCOC$_{alkoxy}$ |
| 570 | 579 | 564 | 30% β$_{asym}$Ph$_{alkoxy}$, 10% δCOC$_{alkoxy}$, 10% δCSC/Ph$_{alkyl}$ |
| 628 | 648 | 631 | 65% β$_{asym}$Ph$_{alkyl}$ |
| 641 | 657 | 640 | 51% γCOS, 16% γPh$_{alkoxy}$ |
| 661 | 669 | 652 | 44% β$_{asym}$Ph$_{alkoxy}$ |
| 670 | 680 | 662 | 58% β$_{sym}$Ph$_{alkyl}$ |
| 723 | 734 | 715 | 83% ρCH$_{2alkoxy}$ |
| 733 | 742 | 723 | 71% ρCH$_{2alkyl}$ |
| 750 | 766 | 746 | 22% ρCH$_{2alkyl}$, 17% γPh$_{alkyl}$ |
| 781 | 792 | 772 | 45% ρCH$_{2alkoxy}$ |
| 803 | 822 | 801 | 51% γPh$_{alkyl}$ |
| 819 | 841 | 819 | 47% β$_{asym}$Ph$_{alkoxy}$ |
| 839 | 863 | 841 | 76% γPh$_{alkoxy}$ |
| 851 | 871 | 848 | 49% γPh$_{alkyl}$, 20% τCH$_{2alkyl}$ |
| 906 | 919 | 895 | 29% β$_{asym}$Ph$_{alkoxy}$, 26% νCS |
| 934 | 993 | 967 | 74% νCC$_{alkoxy}$ |
| 946 | 1001 | 975 | 57% νCC$_{alkoxy}$ |
| 987 | 1028 | 1001 | 69% νCC$_{alkoxy}$ |
| 1015 | 1048 | 1021 | 41% νCO$_{alkoxy}$ |
| 1023 | 1066 | 1038 | 59% νCC$_{alkoxy}$ |
| 1051 | 1076 | 1048 | 67% νCC$_{alkoxy}$ |
| 1095 | 1120 | 1091 | 75% β$_{asym}$Ph$_{alkyl}$ |
| 1119 | 1150 | 1120 | 54% ν$_{sym}$CCC$_{alkoxy}$ |
| 1168 | 1189 | 1158 | 73% β$_{sym}$Ph$_{alkoxy}$ |
| 1212 | 1238 | 1206 | 69% β$_{asym}$Ph$_{alkoxy}$ |
| 1231 | 1266 | 1233 | 59% ωCH$_{2alkyl}$ |
| 1269 | 1296 | 1262 | 66% β$_{asym}$Ph$_{alkoxy}$ |
| 1310 | 1348 | 1313 | 69% β$_{asym}$Ph$_{alkoxy}$ |
| 1323 | 1355 | 1320 | 40% ωCH$_{2alkyl}$, 25% β$_{asym}$Ph$_{alkyl}$ |
| 1353 | 1370 | 1335 | 61% ωCH$_{2alkoxy}$ |
| 1378 | 1395 | 1359 | 55% ωCH$_{2alkyl}$ |
| 1400 | 1432 | 1395 | 65% ωCH$_{2alkoxy}$ |
| 1421 | 1444 | 1407 | 53% β$_{asym}$Ph$_{alkyl}$ |
| 1435 | 1477 | 1439 | 81% δCH$_{2alkyl}$ |
| 1468 | 1523 | 1484 | 81% δCH$_{2alkoxy}$ |
| 1495 | 1535 | 1495 | 67% β$_{asym}$Ph$_{alkyl}$ |
| 1508 | 1549 | 1509 | 63% β$_{asym}$Ph$_{alkoxy}$ |
| 1600 | 1646 | 1603 | 70% β$_{sym}$Ph$_{alkoxy}$ |
| 1661 | 1736 | 1691 | 89% νC=O |
| 2849 | 3017 | 2871 | 74% ν$_{sym}$CH$_{2alkoxy}$ |
| 2872 | 3034 | 2887 | 93% ν$_{sym}$CH$_{2alkoxy}$ |
| 2915 | 3060 | 2912 | 83% ν$_{asym}$CH$_{2alkoxy}$ |
| 2956 | 3078 | 2929 | 75% ν$_{asym}$CH$_{2alkoxy}$ |
| 3041 | 3190 | 3035 | 97% ν$_{asym}$CH/Ph$_{alkoxy}$ |
| 3049 | 3200 | 3045 | 88% ν$_{sym}$CH/Ph$_{alkyl}$ |
| 3060 | 3221 | 3065 | 99% ν$_{sym}$CH/Ph$_{alkoxy}$ |

Table S2. The band assignment of the experimental IR spectra of 11OS5 at the room temperature, based on the DFT calculations (**def2TZVP** basis set, **B3LYP-D3(BJ)** functional). Notations: β – in-plane deformation, γ – out-of-plane deformation, δ – scissoring, ν – stretching, ρ – rocking, τ – twisting, ω – wagging.

| experimental peak position [cm$^{-1}$] | unscaled calculated peak position [cm$^{-1}$] | scaled calculated peak position [cm$^{-1}$] | vibration (contribution ≥ 10%) |
|---|---|---|---|
| 384 | 388 | 378 | 59% δCCC$_{alkyl}$ |
| 401 | 403 | 393 | 25% δCCC$_{alkoxy}$ |
| 416 | 415 | 404 | 80% γPh$_{alkyl}$ |
| 445 | 448 | 437 | 30% δCCC$_{alkoxy}$ |
| 468 | 472 | 460 | 46% δCCO$_{alkoxy}$ |
| 484 | 495 | 482 | 13% γPh$_{alkyl}$, 11% δCCO$_{alkoxy}$ |
| 500 | 505 | 492 | 44% δCCC$_{alkyl}$ |
| 506 | 514 | 502 | 67% γPh$_{alkoxy}$ |
| 523 | 532 | 519 | 45% δCOC$_{alkoxy}$ |
| 570 | 579 | 564 | 25% β$_{asym}$Ph$_{alkoxy}$, 12% δCOC$_{alkoxy}$ |
| 628 | 647 | 631 | 66% β$_{asym}$Ph$_{alkoxy}$ |
| 641 | 657 | 640 | 77% γPh$_{alkoxy}$ |
| 661 | 669 | 652 | 51% β$_{asym}$Ph$_{alkoxy}$ |
| 670 | 680 | 663 | 45% β$_{sym}$Ph$_{alkyl}$ |
| 723 | 735 | 715 | 68% ρCH$_{2alkoxy}$ |
| 733 | 742 | 723 | 67% ρCH$_{2alkyl}$ |
| 750 | 766 | 747 | 36% ρCH$_{2alkyl}$ |
| 781 | 792 | 772 | 46% ρCH$_{2alkoxy}$ |
| 803 | 820 | 799 | 56% γPh$_{alkyl}$ |
| 819 | 841 | 820 | 60% β$_{asym}$Ph$_{alkoxy}$ |
| 839 | 860 | 838 | 82% γPh$_{alkoxy}$ |
| 851 | 869 | 847 | 48% γPh$_{alkyl}$, 10% ρCH$_{2alkyl}$ |
| 906 | 919 | 896 | 29% β$_{asym}$Ph$_{alkoxy}$, 25% νCS |
| 934 | 993 | 968 | 66% νCC$_{alkoxy}$ |
| 946 | 1001 | 976 | 69% νCC$_{alkoxy}$ |
| 987 | 1028 | 1002 | 57% νCC$_{alkoxy}$ |
| 1015 | 1048 | 1021 | 51% νCC$_{alkoxy}$ |
| 1023 | 1067 | 1040 | 73% νCC$_{alkoxy}$ |
| 1051 | 1077 | 1050 | 66% νCC$_{alkoxy}$ |
| 1095 | 1120 | 1091 | 69% β$_{asym}$Ph$_{alkyl}$ |
| 1119 | 1150 | 1121 | 35% ν$_{sym}$CCC$_{alkoxy}$, 12% ωCH$_{2alkoxy}$ |
| 1168 | 1188 | 1158 | 61% β$_{sym}$Ph$_{alkoxy}$ |
| 1212 | 1238 | 1206 | 70% β$_{asym}$Ph$_{alkoxy}$ |
| 1231 | 1265 | 1233 | 56% ωCH$_{2alkyl}$ |
| 1269 | 1295 | 1262 | 49% νCO$_{alkoxy}$, 11% β$_{asym}$Ph$_{alkoxy}$ |
| 1310 | 1348 | 1314 | 61% β$_{asym}$Ph$_{alkoxy}$ |
| 1323 | 1354 | 1319 | 48% ωCH$_{2alkyl}$, 25% β$_{asym}$Ph$_{alkyl}$ |
| 1353 | 1369 | 1334 | 58% ωCH$_{2alkoxy}$ |
| 1378 | 1395 | 1359 | 53% ωCH$_{2alkoxy}$ |
| 1400 | 1430 | 1393 | 70% ωCH$_{2alkoxy}$ |
| 1421 | 1443 | 1406 | 57% β$_{asym}$Ph$_{alkyl}$, 11% ωCH$_{2alkoxy}$ |
| 1435 | 1476 | 1438 | 69% δCH$_{2alkyl}$ |
| 1468 | 1522 | 1483 | 69% δCH$_{2alkoxy}$ |
| 1495 | 1534 | 1495 | 58% β$_{asym}$Ph$_{alkyl}$ |
| 1508 | 1548 | 1508 | 50% β$_{asym}$Ph$_{alkoxy}$, 12% νCO$_{alkoxy}$ |
| 1600 | 1646 | 1604 | 66% β$_{sym}$Ph$_{alkoxy}$ |
| 1661 | 1736 | 1692 | 89% νC=O |
| 2849 | 3017 | 2871 | 85% ν$_{sym}$CH$_{2alkoxy}$ |
| 2872 | 3034 | 2887 | 92% ν$_{sym}$CH$_{2alkoxy}$ |
| 2915 | 3060 | 2912 | 90% ν$_{asym}$CH$_{2alkoxy}$ |
| 2956 | 3077 | 2928 | 90% ν$_{asym}$CH$_{2alkoxy}$ |
| 3041 | 3190 | 3036 | 98% ν$_{asym}$CH/Ph$_{alkoxy}$ |
| 3049 | 3199 | 3044 | 88% ν$_{sym}$CH/Ph$_{alkyl}$ |
| 3060 | 3220 | 3064 | 85% ν$_{sym}$CH/Ph$_{alkoxy}$, 15% ν$_{sym}$CH$_{2alkoxy}$ |

Table S3. The band assignment of the experimental IR spectra of 11OS5 at the room temperature, based on the DFT calculations (**def2SVPP** basis set, **B3LYP-D3(BJ)** functional). Notations: β – in-plane deformation, γ – out-of-plane deformation, δ – scissoring, ν – stretching, ρ – rocking, τ – twisting, ω – wagging.

| experimental peak position [cm$^{-1}$] | unscaled calculated peak position [cm$^{-1}$] | scaled calculated peak position [cm$^{-1}$] | vibration (contribution ≥ 10%) |
|---|---|---|---|
| 384 | 389 | 376 | 65% δCCC$_{alkyl}$ |
| 401 | 406 | 392 | 16% δCCC$_{alkoxy}$ |
| 416 | 421 | 407 | 70% γPh$_{alkyl}$ |
| 445 | 452 | 436 | 17% δCCO$_{alkoxy}$ |
| 468 | 476 | 460 | 54% δCCC$_{alkoxy}$, 10% νCS |
| 484 | 500 | 483 | 11% δCCC$_{alkyl}$ |
| 500 | 507 | 490 | 41% δCCC$_{alkyl}$ |
| 506 | 521 | 503 | 72% γPh$_{alkoxy}$ |
| 523 | 535 | 517 | 41% δCCC$_{alkoxy}$ |
| 570 | 582 | 562 | 16% β$_{asym}$Ph$_{alkoxy}$, 11% δCSC/Ph$_{alkyl}$ |
| 628 | 647 | 625 | 77% β$_{asym}$Ph$_{alkoxy}$ |
| 641 | 656 | 633 | 83% γPh$_{alkoxy}$ |
| 661 | 671 | 648 | 26% β$_{asym}$Ph$_{alkoxy}$ |
| 670 | 680 | 657 | 29% β$_{sym}$Ph$_{alkyl}$, 21% νSC/Ph$_{alkyl}$ |
| 723 | 744 | 718 | 77% ρCH$_{2alkoxy}$ |
| 733 | 751 | 725 | 76% ρCH$_{2alkyl}$ |
| 750 | 772 | 745 | 15% γPh$_{alkyl}$, 11% ρCH$_{2alkyl}$ |
| 781 | 798 | 771 | 49% ρCH$_{2alkoxy}$ |
| 803 | 824 | 796 | 44% γPh$_{alkyl}$, 10% ρCH$_{2alkyl}$ |
| 819 | 846 | 817 | 56% β$_{asym}$Ph$_{alkoxy}$ |
| 839 | 868 | 838 | 75% γPh$_{alkoxy}$ |
| 851 | 873 | 843 | 34% γPh$_{alkyl}$, 17% ρCH$_{2alkyl}$ |
| 906 | 928 | 896 | 37% νCS, 23% β$_{asym}$Ph$_{alkoxy}$ |
| 934 | 1014 | 979 | 68% νCC$_{alkoxy}$ |
| 946 | 1026 | 991 | 74% β$_{asym}$Ph$_{alkoxy}$ |
| 987 | 1035 | 999 | 65% β$_{asym}$Ph$_{alkyl}$ |
| 1015 | 1084 | 1041 | 68% νCC$_{alkoxy}$ |
| 1023 | 1090 | 1047 | 69% νCC$_{alkoxy}$ |
| 1051 | 1118 | 1073 | 40% ν$_{sym}$CCC$_{alkoxy}$ |
| 1095 | 1136 | 1091 | 89% β$_{sym}$Ph$_{alkoxy}$ |
| 1119 | 1154 | 1108 | 47% ν$_{sym}$CCC$_{alkoxy}$ |
| 1168 | 1181 | 1134 | 87% β$_{sym}$Ph$_{alkoxy}$ |
| 1212 | 1248 | 1198 | 59% β$_{asym}$Ph$_{alkoxy}$ |
| 1231 | 1301 | 1249 | 60% ωCH$_{2alkoxy}$ |
| 1269 | 1325 | 1272 | 59% β$_{asym}$Ph$_{alkoxy}$ |
| 1310 | 1377 | 1322 | 52% β$_{asym}$Ph$_{alkoxy}$ |
| 1323 | 1399 | 1343 | 58% ωCH$_{2alkoxy}$ |
| 1353 | 1401 | 1345 | 60% ωCH$_{2alkyl}$ |
| 1378 | 1419 | 1362 | 50% ωCH$_{2alkoxy}$ |
| 1400 | 1446 | 1388 | 46% ωCH$_{2alkoxy}$ |
| 1421 | 1449 | 1391 | 52% β$_{asym}$Ph$_{alkyl}$, 11% ωCH$_{2alkyl}$ |
| 1435 | 1482 | 1423 | 98% δCH$_{3alkyl}$ |
| 1468 | 1512 | 1452 | 74% δCH$_{2alkoxy}$ |
| 1495 | 1538 | 1477 | 76% β$_{asym}$Ph$_{alkyl}$ |
| 1508 | 1558 | 1496 | 81% β$_{asym}$Ph$_{alkoxy}$ |
| 1600 | 1671 | 1604 | 72% β$_{sym}$Ph$_{alkoxy}$ |
| 1661 | 1787 | 1716 | 91% νC=O |
| 2849 | 3006 | 2862 | 82% ν$_{sym}$CH$_{2alkoxy}$ |
| 2872 | 3025 | 2880 | 96% ν$_{sym}$CH$_{2alkoxy}$ |
| 2915 | 3060 | 2914 | 89% ν$_{asym}$CH$_{2alkoxy}$ |
| 2956 | 3079 | 2932 | 80% ν$_{asym}$CH$_{2alkoxy}$ |
| 3041 | 3191 | 3039 | 95% ν$_{asym}$CH/Ph$_{alkoxy}$ |
| 3049 | 3203 | 3050 | 95% ν$_{sym}$CH/Ph$_{alkyl}$ |
| 3060 | 3219 | 3065 | 90% ν$_{sym}$CH/Ph$_{alkoxy}$ |

Table S4. The band assignment of the experimental IR spectra of 11OS5 at the room temperature, based on the DFT calculations (**def2SVP** basis set, **B3LYP-D3(BJ)** functional). Notations: β – in-plane deformation, γ – out-of-plane deformation, δ – scissoring, ν – stretching, ρ – rocking, τ – twisting, ω – wagging.

| experimental peak position [cm$^{-1}$] | unscaled calculated peak position [cm$^{-1}$] | scaled calculated peak position [cm$^{-1}$] | vibration (contribution ≥ 10%) |
|---|---|---|---|
| 384 | 388 | 376 | 57% δCCC$_{alkyl}$ |
| 401 | 406 | 393 | 44% δCCC$_{alkoxy}$ |
| 416 | 421 | 408 | 67% γPh$_{alkyl}$ |
| 445 | 451 | 437 | 20% δCCC$_{alkoxy}$ |
| 468 | 475 | 460 | 47% δCCC$_{alkoxy}$ |
| 484 | 499 | 483 | 21% γPh$_{alkyl}$ |
| 500 | 507 | 491 | 32% δCCC$_{alkyl}$, 18% γPh$_{alkyl}$ |
| 506 | 520 | 504 | 74% γPh$_{alkoxy}$ |
| 523 | 535 | 518 | 45% δCCO$_{alkoxy}$, 11% γPh$_{alkyl}$ |
| 570 | 581 | 563 | 11% β$_{asym}$Ph$_{alkoxy}$, 10% δCCO$_{alkoxy}$ |
| 628 | 646 | 626 | 47% β$_{asym}$Ph$_{alkoxy}$ |
| 641 | 655 | 634 | 60% γCOS, 14% γPh$_{alkoxy}$ |
| 661 | 671 | 650 | 37% β$_{asym}$Ph$_{alkoxy}$ |
| 670 | 680 | 659 | 60% β$_{sym}$Ph$_{alkyl}$ |
| 723 | 741 | 718 | 69% ρCH$_{2alkoxy}$ |
| 733 | 745 | 722 | 69% ρCH$_{2alkyl}$ |
| 750 | 767 | 743 | 18% ρCH$_{2alkyl}$, 15% γPh$_{alkyl}$ |
| 781 | 790 | 765 | 61% ρCH$_{2alkoxy}$ |
| 803 | 820 | 794 | 51% ρCH$_{2alkyl}$ |
| 819 | 846 | 819 | 57% β$_{asym}$Ph$_{alkoxy}$ |
| 839 | 866 | 839 | 62% γPh$_{alkoxy}$ |
| 851 | 867 | 840 | 35% γPh$_{alkyl}$, 11% γPh$_{alkoxy}$ |
| 906 | 928 | 899 | 40% νCS, 25% β$_{asym}$Ph$_{alkoxy}$ |
| 934 | 1012 | 980 | 75% νCC$_{alkoxy}$ |
| 946 | 1022 | 990 | 73% β$_{asym}$Ph$_{alkoxy}$ |
| 987 | 1032 | 1000 | 74% β$_{asym}$Ph$_{alkoxy}$ |
| 1015 | 1081 | 1043 | 55% νCC$_{alkoxy}$ |
| 1023 | 1088 | 1049 | 57% νCC$_{alkoxy}$ |
| 1051 | 1113 | 1073 | 36% νCC$_{alkoxy}$ |
| 1095 | 1132 | 1092 | 89% β$_{asym}$Ph$_{alkoxy}$ |
| 1119 | 1150 | 1109 | 72% ν$_{sym}$CCC$_{alkoxy}$ |
| 1168 | 1177 | 1135 | 88% β$_{sym}$Ph$_{alkoxy}$ |
| 1212 | 1246 | 1202 | 66% β$_{asym}$Ph$_{alkoxy}$ |
| 1231 | 1287 | 1241 | 60% ωCH$_{2alkoxy}$ |
| 1269 | 1320 | 1273 | 26% ωCH$_{2alkoxy}$, 21% β$_{asym}$Ph$_{alkoxy}$ |
| 1269 | 1326 | 1279 | 27% ωCH$_{2alkoxy}$, 20% β$_{asym}$Ph$_{alkoxy}$ |
| 1310 | 1374 | 1325 | 58% β$_{asym}$Ph$_{alkoxy}$ |
| 1323 | 1386 | 1337 | 61% ωCH$_{2alkoxy}$ |
| 1353 | 1389 | 1341 | 50% ωCH$_{2alkyl}$ |
| 1378 | 1407 | 1357 | 66% ωCH$_{2alkoxy}$, 10% νCC$_{alkoxy}$ |
| 1400 | 1435 | 1384 | 51% ωCH$_{2alkoxy}$ |
| 1421 | 1443 | 1392 | 52% β$_{asym}$Ph$_{alkyl}$, 15% ωCH$_{2alkyl}$ |
| 1435 | 1471 | 1419 | 98% δCH$_{3alkyl}$ |
| 1468 | 1500 | 1447 | 60% δCH$_{2alkoxy}$ |
| 1495 | 1530 | 1476 | 80% β$_{asym}$Ph$_{alkyl}$ |
| 1508 | 1552 | 1497 | 80% β$_{asym}$Ph$_{alkoxy}$ |
| 1600 | 1667 | 1608 | 75% β$_{sym}$Ph$_{alkoxy}$ |
| 1661 | 1787 | 1724 | 90% νC=O |
| 2849 | 3018 | 2866 | 87% ν$_{sym}$CH$_{2alkoxy}$ |
| 2872 | 3038 | 2885 | 96% ν$_{sym}$CH$_{2alkoxy}$ |
| 2915 | 3071 | 2916 | 91% ν$_{asym}$CH$_{2alkoxy}$ |
| 2956 | 3090 | 2934 | 84% ν$_{asym}$CH$_{2alkoxy}$ |
| 3041 | 3196 | 3035 | 95% ν$_{asym}$CH/Ph$_{alkoxy}$ |
| 3049 | 3207 | 3045 | 97% ν$_{sym}$CH/Ph$_{alkyl}$ |
| 3060 | 3224 | 3061 | 99% ν$_{sym}$CH/Ph$_{alkoxy}$ |

Table S5. The band assignment of the experimental IR spectra of 11OS5 at the room temperature, based on the DFT calculations (**def2SVP** basis set, **BLYP-D3(BJ)** functional). Notations: β – in-plane deformation, γ – out-of-plane deformation, δ – scissoring, ν – stretching, ρ – rocking, τ – twisting, ω – wagging.

| experimental peak position [cm$^{-1}$] | unscaled calculated peak position [cm$^{-1}$] | scaled calculated peak position [cm$^{-1}$] | vibration (contribution ≥ 10%) |
|---|---|---|---|
| 384 | 378 | 380 | 61% δCCC$_{alkyl}$ |
| 401 | 392 | 394 | 34% δCCC$_{alkoxy}$, 10% νCS |
| 416 | 423 | 425 | 33% ν$_{asym}$CSC/Ph$_{alkyl}$, 10% δCCC$_{alkyl}$ |
| 445 | 438 | 440 | 12% δCCO$_{alkoxy}$ |
| 468 | 460 | 462 | 47% δCOC/Ph$_{alkoxy}$ |
| 484 | 485 | 487 | 11% δCOC$_{alkoxy}$ |
| 500 | 493 | 495 | 40% δCCC$_{alkyl}$ |
| 506 | 504 | 506 | 30% δCCC$_{alkoxy}$ |
| 523 | 519 | 522 | 49% δCOC$_{alkoxy}$ |
| 570 | 560 | 563 | 23% δCSC/Ph$_{alkyl}$, 20% δCCC$_{alkyl}$, 12% δCOC$_{alkoxy}$ |
| 628 | 627 | 630 | 58% β$_{asym}$Ph$_{alkoxy}$, 11% γCOS |
| 641 | 629 | 632 | 52% γCOS, 12% γPh$_{alkoxy}$ |
| 661 | 647 | 650 | 42% β$_{asym}$Ph$_{alkoxy}$ |
| 670 | 657 | 660 | 31% νSC/Ph$_{alkyl}$, 18% β$_{sym}$Ph$_{alkyl}$ |
| 723 | 730 | 734 | 75% ρCH$_{2alkyl}$ |
| 733 | 737 | 741 | 74% γPh$_{alkoxy}$ |
| 750 | 746 | 750 | 36% ρCH$_{2alkyl}$, 27% γPh$_{alkyl}$ |
| 781 | 772 | 776 | 37% ρCH$_{2alkoxy}$ |
| 803 | 796 | 800 | 58% ρCH$_{2alkyl}$ |
| 819 | 814 | 818 | 34% β$_{asym}$Ph$_{alkoxy}$ |
| 839 | 837 | 841 | 78% γPh$_{alkoxy}$ |
| 851 | 842 | 846 | 40% ρCH$_{2alkyl}$, 11% γPh$_{alkyl}$ |
| 906 | 888 | 892 | 36% νCS, 33% β$_{asym}$Ph$_{alkoxy}$ |
| 934 | 956 | 961 | 88% γPh$_{alkoxy}$ |
| 946 | 962 | 967 | 92% γPh$_{alkyl}$ |
| 987 | 974 | 979 | 50% νCC$_{alkoxy}$ |
| 1015 | 1002 | 1005 | 49% νCC$_{alkoxy}$ |
| 1023 | 1013 | 1016 | 40% νCO$_{alkoxy}$ |
| 1051 | 1047 | 1051 | 51% νCC$_{alkoxy}$ |
| 1095 | 1078 | 1082 | 75% β$_{asym}$Ph$_{alkyl}$ |
| 1119 | 1102 | 1106 | 88% β$_{sym}$Ph$_{alkoxy}$ |
| 1168 | 1144 | 1148 | 73% β$_{sym}$Ph$_{alkoxy}$ |
| 1212 | 1202 | 1206 | 66% β$_{asym}$Ph$_{alkoxy}$ |
| 1231 | 1219 | 1223 | 53% ωCH$_{2alkyl}$ |
| 1269 | 1265 | 1269 | 55% νCO$_{alkoxy}$, 14% β$_{asym}$Ph$_{alkoxy}$ |
| 1310 | 1340 | 1345 | 65% β$_{sym}$Ph$_{alkoxy}$ |
| 1323 | 1345 | 1350 | 33% ωCH$_{2alkoxy}$, 13% β$_{sym}$Ph$_{alkoxy}$ |
| 1353 | 1350 | 1355 | 57% ωCH$_{2alkyl}$, 14% β$_{sym}$Ph$_{alkyl}$ |
| 1378 | 1359 | 1364 | 67% ωCH$_{2alkoxy}$ |
| 1400 | 1384 | 1389 | 52% ωCH$_{2alkoxy}$ |
| 1421 | 1396 | 1401 | 60% β$_{asym}$Ph$_{alkyl}$, 13% ωCH$_{2alkoxy}$ |
| 1435 | 1432 | 1437 | 86% δCH$_{3alkyl}$ |
| 1468 | 1457 | 1462 | 69% δCH$_{2alkoxy}$ |
| 1495 | 1478 | 1483 | 77% β$_{asym}$Ph$_{alkyl}$ |
| 1508 | 1494 | 1499 | 57% β$_{asym}$Ph$_{alkoxy}$ |
| 1600 | 1603 | 1608 | 67% β$_{sym}$Ph$_{alkoxy}$ |
| 1661 | 1707 | 1713 | 90% νC=O |
| 2849 | 2939 | 2866 | 79% ν$_{sym}$CH$_{2alkoxy}$ |
| 2872 | 2958 | 2885 | 96% ν$_{sym}$CH$_{2alkoxy}$ |
| 2915 | 2988 | 2914 | 91% ν$_{asym}$CH$_{2alkoxy}$ |
| 2956 | 3006 | 2932 | 83% ν$_{asym}$CH$_{2alkoxy}$ |
| 3041 | 3113 | 3036 | 96% ν$_{asym}$CH/Ph$_{alkoxy}$ |
| 3049 | 3124 | 3047 | 99% ν$_{sym}$CH/Ph$_{alkyl}$ |
| 3060 | 3141 | 3063 | 100% ν$_{sym}$CH/Ph$_{alkoxy}$ |

Table S6. The band assignment of the experimental IR spectra of 11OS5 at the room temperature, based on the DFT calculations (**6311+Gdp** basis set, **B3LYP-D3(BJ)** functional). Notations: β – in-plane deformation, γ – out-of-plane deformation, δ – scissoring, ν – stretching, ρ – rocking, τ – twisting, ω – wagging.

| experimental peak position [cm$^{-1}$] | unscaled calculated peak position [cm$^{-1}$] | scaled calculated peak position [cm$^{-1}$] | vibration (contribution ≥ 10%) |
|---|---|---|---|
| 384 | 381 | 373 | 35% δCCC$_{alkoxy}$, 13% δOCC/Ph$_{alkoxy}$ |
| 401 | 402 | 393 | 25% δCCC$_{alkoxy}$ |
| 416 | 422 | 413 | 28 % νSC, 25% δCCC$_{alkyl}$ |
| 445 | 447 | 437 | 17% δCCC$_{alkoxy}$ |
| 468 | 473 | 463 | 13% δOCC/Ph$_{alkoxy}$ |
| 484 | 507 | 496 | 32% δCCC$_{alkoxy}$ |
| 500 | 508 | 497 | 77% γPh$_{alkoxy}$ |
| 506 | 519 | 508 | 38% δCCC$_{alkoxy}$ |
| 523 | 546 | 534 | 17% γPh$_{alkyl}$ |
| 570 | 585 | 572 | 14% γPh$_{alkyl}$ |
| 628 | 645 | 631 | 74% β$_{asym}$Ph$_{alkoxy}$ |
| 641 | 646 | 632 | 80% γPh$_{alkoxy}$ |
| 661 | 665 | 650 | 40% β$_{asym}$Ph$_{alkoxy}$ |
| 670 | 673 | 658 | 30% β$_{sym}$Ph$_{alkyl}$ |
| 723 | 733 | 717 | 65% ρCH$_{2alkyl}$ |
| 733 | 742 | 726 | 73% ρCH$_{2alkoxy}$ |
| 750 | 757 | 740 | 75% γPh$_{alkyl}$ |
| 781 | 791 | 774 | 53% ρCH$_{2alkoxy}$ |
| 803 | 824 | 806 | 49% γPh$_{alkyl}$ |
| 819 | 838 | 820 | 29% νCC$_{alkoxy}$, 17% νCO$_{alkoxy}$, 13% β$_{asym}$Ph$_{alkoxy}$ |
| 839 | 850 | 831 | 68% γPh$_{alkoxy}$ |
| 851 | 850 | 831 | 49% γPh$_{alkyl}$ |
| 906 | 914 | 894 | 24% νCS, 22% β$_{asym}$Ph$_{alkoxy}$ |
| 934 | 985 | 963 | 86% γPh$_{alkoxy}$ |
| 946 | 999 | 977 | 55% νCC$_{alkoxy}$ |
| 987 | 1025 | 1002 | 41% νCC$_{alkoxy}$, 21% β$_{asym}$Ph$_{alkoxy}$ |
| 1015 | 1044 | 1018 | 49% νCO$_{alkoxy}$ |
| 1023 | 1064 | 1037 | 70% νCC$_{alkoxy}$ |
| 1051 | 1075 | 1048 | 63% νCC$_{alkoxy}$ |
| 1095 | 1114 | 1086 | 78% β$_{asym}$Ph$_{alkyl}$ |
| 1119 | 1147 | 1118 | 58% ν$_{sym}$CCC$_{alkoxy}$ |
| 1168 | 1186 | 1156 | 62% β$_{sym}$Ph$_{alkoxy}$ |
| 1212 | 1236 | 1205 | 71% β$_{asym}$Ph$_{alkoxy}$ |
| 1231 | 1266 | 1234 | 67% ωCH$_{2alkoxy}$ |
| 1269 | 1290 | 1258 | 46% νCO$_{alkoxy}$, 23% β$_{asym}$Ph$_{alkoxy}$ |
| 1310 | 1347 | 1313 | 72% β$_{asym}$Ph$_{alkoxy}$ |
| 1323 | 1348 | 1314 | 52% β$_{asym}$Ph$_{alkyl}$, 14% τCH$_{2alkyl}$ |
| 1353 | 1365 | 1331 | 65% ωCH$_{2alkoxy}$ |
| 1378 | 1413 | 1377 | 83% δCH$_{3alkoxy}$ |
| 1400 | 1430 | 1394 | 57% ωCH$_{2alkoxy}$ |
| 1421 | 1435 | 1399 | 37% τCH$_{2alkyl}$, 34% β$_{asym}$Ph$_{alkyl}$ |
| 1435 | 1500 | 1462 | 99% δCH$_{3alkyl}$ |
| 1468 | 1521 | 1483 | 68% δCH$_{2alkoxy}$ |
| 1495 | 1525 | 1487 | 73% β$_{asym}$Ph$_{alkyl}$ |
| 1508 | 1542 | 1503 | 47% β$_{asym}$Ph$_{alkoxy}$ |
| 1600 | 1643 | 1602 | 71% β$_{sym}$Ph$_{alkoxy}$ |
| 1661 | 1742 | 1698 | 90% νC=O |
| 2849 | 3012 | 2870 | 86% ν$_{sym}$CH$_{2alkoxy}$ |
| 2872 | 3032 | 2889 | 94% ν$_{sym}$CH$_{2alkoxy}$ |
| 2915 | 3057 | 2913 | 82% ν$_{asym}$CH$_{2alkoxy}$ |
| 2956 | 3079 | 2934 | 84% ν$_{asym}$CH$_{2alkyl}$ |
| 3041 | 3186 | 3036 | 96% ν$_{asym}$CH/Ph$_{alkoxy}$ |
| 3049 | 3190 | 3040 | 94% ν$_{asym}$CH/Ph$_{alkyl}$ |
| 3060 | 3214 | 3062 | 100% ν$_{sym}$CH/Ph$_{alkoxy}$ |

Table S7. The band assignment of the experimental IR spectra of 11OS5 at the room temperature, based on the DFT calculations (**631+Gd** basis set, **B3LYP-D3(BJ)** functional). Notations: β – in-plane deformation, γ – out-of-plane deformation, δ – scissoring, ν – stretching, ρ – rocking, τ – twisting, ω – wagging.

| experimental peak position [cm$^{-1}$] | unscaled calculated peak position [cm$^{-1}$] | scaled calculated peak position [cm$^{-1}$] | vibration (contribution ≥ 10%) |
|---|---|---|---|
| 384 | 387 | 378 | 55% δCCC$_{alkyl}$ |
| 401 | 404 | 395 | 14% δCCC$_{alkoxy}$ |
| 416 | 433 | 423 | 35% νCS |
| 445 | 448 | 438 | 31% δCCC$_{alkoxy}$ |
| 468 | 470 | 459 | 26% δCCC$_{alkoxy}$ |
| 484 | 492 | 481 | 27% γPh$_{alkyl}$ |
| 500 | 503 | 491 | 43% δCCC$_{alkyl}$ |
| 506 | 508 | 496 | 84% γPh$_{alkoxy}$ |
| 523 | 531 | 519 | 43% δCCO$_{alkoxy}$ |
| 570 | 577 | 564 | 21% δCCC$_{alkoxy}$, 13% β$_{asym}$Ph$_{alkoxy}$, 10% νCS |
| 628 | 645 | 630 | 52% β$_{asym}$Ph$_{alkoxy}$, 14% δCCC$_{alkoxy}$ |
| 641 | 646 | 631 | 75% γCOS, 13% γPh$_{alkoxy}$ |
| 661 | 666 | 650 | 39% β$_{asym}$Ph$_{alkoxy}$ |
| 670 | 677 | 661 | 29% νSC/Ph$_{alkyl}$, 20% β$_{sym}$Ph$_{alkyl}$ |
| 723 | 736 | 719 | 73% ρCH$_{2alkoxy}$ |
| 733 | 743 | 726 | 74% ρCH$_{2alkyl}$ |
| 750 | 765 | 747 | 32% ρCH$_{2alkyl}$, 23% γPh$_{alkyl}$ |
| 781 | 795 | 776 | 51% ρCH$_{2alkoxy}$ |
| 803 | 818 | 799 | 67% γPh$_{alkyl}$ |
| 819 | 839 | 819 | 29% β$_{asym}$Ph$_{alkoxy}$, 18% νCO$_{alkoxy}$ |
| 839 | 850 | 830 | 46% γPh$_{alkoxy}$ |
| 851 | 870 | 850 | 36% γPh$_{alkyl}$, 21% ρCH$_{2alkyl}$ |
| 906 | 919 | 898 | 35% νCS, 26% β$_{asym}$Ph$_{alkoxy}$ |
| 934 | 982 | 959 | 80% γPh$_{alkoxy}$ |
| 946 | 1004 | 981 | 64% νCC$_{alkoxy}$ |
| 987 | 1029 | 1005 | 62% β$_{asym}$Ph$_{alkoxy}$, 18% νCC$_{alkoxy}$ |
| 1015 | 1051 | 1015 | 45% νCC$_{alkoxy}$ |
| 1023 | 1071 | 1035 | 68% νCC$_{alkoxy}$ |
| 1051 | 1081 | 1044 | 73% νCC$_{alkoxy}$ |
| 1095 | 1122 | 1084 | 70% νSC/Ph$_{alkoxy}$ |
| 1119 | 1154 | 1115 | 68% β$_{sym}$Ph$_{alkoxy}$ |
| 1168 | 1198 | 1157 | 90% β$_{sym}$Ph$_{alkoxy}$ |
| 1212 | 1245 | 1203 | 71% β$_{asym}$Ph$_{alkoxy}$ |
| 1231 | 1271 | 1228 | 53% ωCH$_{2alkoxy}$ |
| 1269 | 1303 | 1259 | 52% ν$_{asym}$COC$_{alkoxy}$, 12% β$_{asym}$Ph$_{alkoxy}$ |
| 1310 | 1363 | 1317 | 36% β$_{sym}$Ph$_{alkoxy}$, 14% ωCH$_{2alkyl}$ |
| 1323 | 1364 | 1318 | 24% ωCH$_{2alkyl}$, 20% β$_{asym}$Ph$_{alkoxy}$ |
| 1353 | 1377 | 1330 | 51% ωCH$_{2alkoxy}$ |
| 1378 | 1434 | 1385 | 83% ωCH$_{3alkyl}$ |
| 1400 | 1443 | 1394 | 57% ωCH$_{2alkoxy}$ |
| 1421 | 1450 | 1401 | 52% β$_{asym}$Ph$_{alkyl}$, 14% ωCH$_{2alkyl}$ |
| 1435 | 1522 | 1470 | 82% δCH$_{3alkyl}$, 10% τCH$_{2alkoxy}$ |
| 1468 | 1539 | 1487 | 67% δCH$_{2alkoxy}$ |
| 1495 | 1541 | 1489 | 63% β$_{asym}$Ph$_{alkyl}$ |
| 1508 | 1557 | 1504 | 42% β$_{asym}$Ph$_{alkoxy}$ |
| 1600 | 1655 | 1599 | 73% β$_{sym}$Ph$_{alkoxy}$ |
| 1661 | 1746 | 1687 | 89% νC=O |
| 2849 | 3025 | 2866 | 81% ν$_{sym}$CH$_{2alkoxy}$ |
| 2872 | 3046 | 2886 | 95% ν$_{sym}$CH$_{2alkoxy}$ |
| 2915 | 3071 | 2910 | 82% ν$_{asym}$CH$_{2alkoxy}$ |
| 2956 | 3093 | 2930 | 90% ν$_{asym}$CH$_{2alkoxy}$ |
| 3041 | 3209 | 3040 | 93% ν$_{asym}$CH/Ph$_{alkyl}$ |
| 3049 | 3216 | 3047 | 94% ν$_{asym}$CH/Ph$_{alkyl}$ |
| 3060 | 3233 | 3063 | 91% ν$_{sym}$CH/Ph$_{alkoxy}$ |

Table S8. The band assignment of the experimental IR spectra of 11OS5 at the room temperature, based on the DFT calculations (**631+Gd** basis set, **BLYP-D3(BJ)** functional). Notations: β – in-plane deformation, γ – out-of-plane deformation, δ – scissoring, ν – stretching, ρ – rocking, τ – twisting, ω – wagging.

| experimental peak position [cm$^{-1}$] | unscaled calculated peak position [cm$^{-1}$] | scaled calculated peak position [cm$^{-1}$] | vibration (contribution ≥ 10%) |
|---|---|---|---|
| 384 | 388 | 393 | 16% δCCO$_{alkoxy}$, 10% νCS |
| 401 | 410 | 416 | 25% ν$_{asym}$CSC/Ph$_{alkyl}$, 27% δCCC$_{alkyl}$ |
| 416 | 420 | 426 | 11% δCCC$_{alkyl}$ |
| 445 | 442 | 448 | 29% δCCC$_{alkoxy}$ |
| 468 | 462 | 468 | 25% δCCC$_{alkoxy}$ |
| 484 | 489 | 496 | 80% γPh$_{alkoxy}$ |
| 500 | 497 | 504 | 50% δCCC$_{alkyl}$ |
| 506 | 507 | 514 | 40% δCCC$_{alkyl}$ |
| 523 | 527 | 534 | 35% γPh$_{alkyl}$, 13% δCCC$_{alkoxy}$ |
| 570 | 554 | 562 | 15% β$_{asym}$Ph$_{alkoxy}$ |
| 628 | 618 | 627 | 85% γPh$_{alkoxy}$ |
| 641 | 625 | 634 | 63% β$_{asym}$Ph$_{alkoxy}$ |
| 661 | 642 | 651 | 56% β$_{asym}$Ph$_{alkoxy}$ |
| 670 | 652 | 661 | 49% ν$_{sym}$CSC/Ph$_{alkyl}$, 11% β$_{sym}$Ph$_{alkyl}$ |
| 723 | 712 | 722 | 80% γPh$_{alkyl}$ |
| 733 | 721 | 731 | 83% ρCH$_{2alkoxy}$ |
| 750 | 749 | 759 | 72% ρCH$_{2alkyl}$ |
| 781 | 776 | 787 | 32% ρCH$_{2alkoxy}$ |
| 803 | 796 | 807 | 55% γPh$_{alkyl}$ |
| 819 | 805 | 816 | 46% β$_{asym}$Ph$_{alkoxy}$ |
| 839 | 812 | 823 | 54% γPh$_{alkoxy}$, 12% γPh$_{alkyl}$ |
| 851 | 818 | 829 | 42% γPh$_{alkyl}$ |
| 906 | 877 | 889 | 61% β$_{asym}$Ph$_{alkoxy}$ |
| 934 | 931 | 944 | 44% τCH$_{2alkoxy}$ |
| 946 | 958 | 971 | 82% νCC$_{alkoxy}$ |
| 987 | 963 | 976 | 64% νCO$_{alkoxy}$ |
| 1015 | 986 | 990 | 74% νCO$_{alkoxy}$ |
| 1023 | 999 | 1003 | 64% νCC$_{alkoxy}$ |
| 1051 | 1027 | 1031 | 72% νCC$_{alkoxy}$ |
| 1095 | 1075 | 1079 | 70% β$_{asym}$Ph$_{alkyl}$ |
| 1119 | 1099 | 1103 | 48% νCC$_{alkyl}$, 21% ωCH$_{2alkyl}$ |
| 1168 | 1161 | 1166 | 76% β$_{sym}$Ph$_{alkoxy}$ |
| 1212 | 1203 | 1208 | 62% β$_{asym}$Ph$_{alkoxy}$ |
| 1231 | 1206 | 1211 | 64% ωCH$_{2alkoxy}$ |
| 1269 | 1245 | 1250 | 56% β$_{asym}$Ph$_{alkoxy}$ |
| 1310 | 1331 | 1336 | 53% β$_{asym}$Ph$_{alkyl}$, 14% τCH$_{2alkyl}$ |
| 1323 | 1331 | 1336 | 39% β$_{sym}$Ph$_{alkoxy}$, 14% νCS, 11% ωCH$_{2alkoxy}$ |
| 1353 | 1337 | 1342 | 58% ωCH$_{2alkoxy}$ |
| 1378 | 1366 | 1372 | 69% ωCH$_{2alkoxy}$ |
| 1400 | 1392 | 1398 | 63% ωCH$_{2alkoxy}$ |
| 1421 | 1401 | 1407 | 64% β$_{asym}$Ph$_{alkyl}$ |
| 1435 | 1483 | 1489 | 71% δCH$_{2alkoxy}$ |
| 1468 | 1492 | 1498 | 61% δCH$_{2alkoxy}$ |
| 1495 | 1496 | 1502 | 68% δCH$_{2alkoxy}$ |
| 1508 | 1499 | 1505 | 69% β$_{asym}$Ph$_{alkoxy}$ |
| 1600 | 1590 | 1596 | 69% β$_{sym}$Ph$_{alkoxy}$ |
| 1661 | 1667 | 1674 | 87% νC=O |
| 2849 | 2942 | 2864 | 80% ν$_{sym}$CH$_{2alkoxy}$ |
| 2872 | 2964 | 2886 | 92% ν$_{sym}$CH$_{2alkoxy}$ |
| 2915 | 2987 | 2908 | 81% ν$_{asym}$CH$_{2alkoxy}$ |
| 2956 | 3009 | 2929 | 83% ν$_{asym}$CH$_{2alkoxy}$ |
| 3041 | 3121 | 3038 | 84% ν$_{sym}$CH/Ph$_{alkyl}$ |
| 3049 | 3135 | 3052 | 99% ν$_{sym}$CH/Ph$_{alkoxy}$ |
| 3060 | 3148 | 3065 | 90% ν$_{sym}$CH/Ph$_{alkoxy}$ |

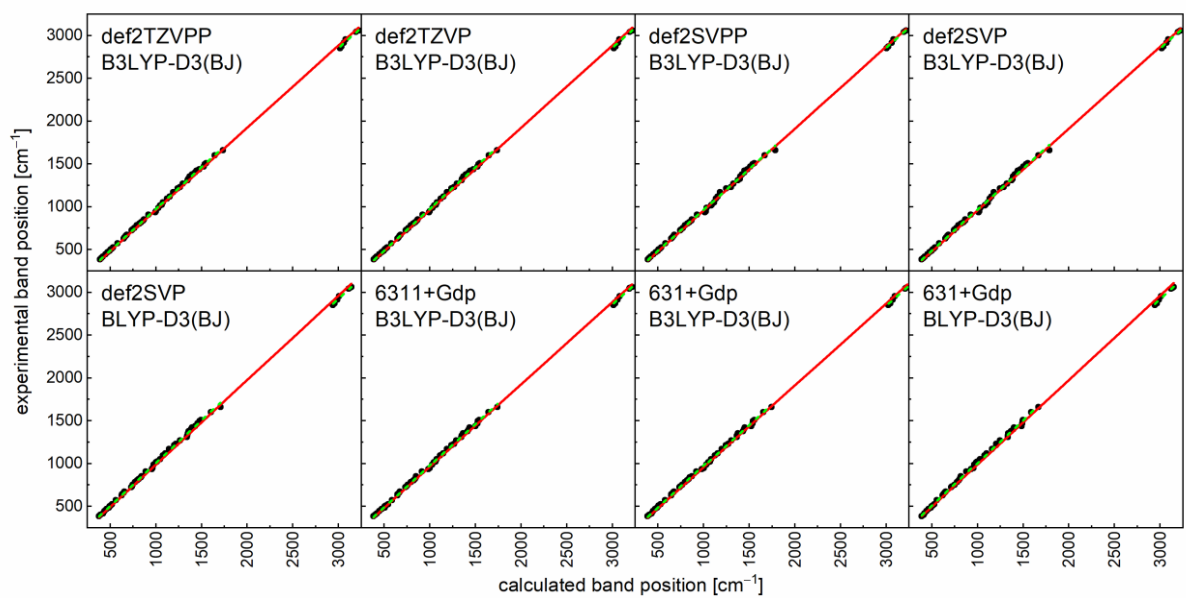

Figure S1. Experimental band positions in the IR spectrum of 11OS5 in the crystal phase in the room temperature vs. calculated band positions at different levels of theory. Solid and dashed lines indicate linear fits with intercept fixed to zero in the full spectral range and separate in the <1000 cm$^{-1}$, 1000-2000 cm$^{-1}$, >2000 cm$^{-1}$ ranges, respectively.

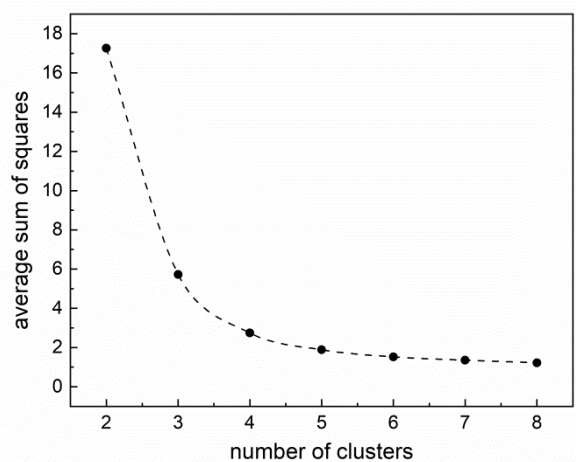

Figure S2. Determination of an optimal number of clusters in the k-means cluster analysis by the "elbow method".